\documentclass[11pt]{article}

\usepackage{fullpage}
\usepackage{amssymb}

\newcounter{subequation}
\newenvironment{subequation}%
{\addtocounter{equation}{-1}%
\stepcounter{subequation}%
\begin{equation}}%
{\end{equation}%
}

\newcommand{\beq}{\begin{equation}}
\newcommand{\eeq}{\end{equation}}
\newcommand{\bseq}{\begin{subequation}}
\newcommand{\eseq}{\end{subequation}}
\newcommand{\bea}{\begin{eqnarray}}
\newcommand{\eea}{\end{eqnarray}}
\newcommand{\refeq}[1]{(\ref{#1})}

\newcommand{\abs}[1]{\left|{#1}\right|}

\newcommand{\pr}{\prime}

\newcommand{\vareps}{\varepsilon}

\newcommand{\sign}{{\rm sign\, }}

\newcount\thmno \thmno=0

\def\proclaim#1{\noindent{\bf #1}}
\def\thmlbl#1{\global\advance\thmno by 1
  \global\edef#1{{\number\thmno}}
 \number\thmno}

\newcommand{\bbN}{\mathbb N}
\newcommand{\bbR}{\mathbb R}
\newcommand{\bbS}{\mathbb S}
\begin{document}

\baselineskip 16pt 

\title{ STATISTICAL MECHANICS APPROACH TO SOME \\
 		PROBLEMS IN CONFORMAL GEOMETRY } 

\author{MICHAEL K.-H. KIESSLING  \smallskip\\
	Department of Mathematics, Rutgers University \\
	110 Frelinghuysen Rd., Piscataway, N.J. 08854\\ 
	email: miki@math.rutgers.edu \\ \\ \\
\it To Joel Louis Lebowitz on occasion of his 70$^{th}$
	birthday. With affection.}

\date{To appear in Physica A, May 2000}

\maketitle

\begin{abstract}
\noindent 
A weak law of large numbers is established for a sequence of systems
of $N$ classical point particles with logarithmic pair potential 
in $\bbR^n$, or $\bbS^n$, $n\in \bbN$, which are distributed 
according to the configurational microcanonical measure $\delta(E-H)$, 
or rather some regularization thereof, where $H$ is the configurational 
Hamiltonian and $E$ the configurational energy. 
When $N\to\infty$ with non-extensive energy scaling 
$E=N^2 \vareps$, the particle positions become i.i.d. according 
to a self-consistent Boltzmann distribution, respectively a 
superposition of such distributions. The self-consistency condition 
in $n$ dimensions is some nonlinear elliptic PDE of order $n$ 
(pseudo-PDE if $n$ is odd) with an exponential nonlinearity. 
When $n=2$, this PDE is known in statistical 
mechanics as Poisson-Boltzmann equation, with applications to point 
vortices, 2D Coulomb and magnetized plasmas and gravitational systems.
It is then also known in conformal differential geometry, 
where it is the central equation in Nirenberg's problem of 
prescribed Gaussian curvature. For constant Gauss curvature  it
becomes Liouville's equation, which also appears in two-dimensional 
so-called quantum Liouville gravity. The PDE for $n=4$ is 
Paneitz' equation, and while it is not known in 
statistical mechanics, it originated from a study of the 
conformal invariance of Maxwell's electromagnetism and 
has made its appearance in some recent model of four-dimensional
quantum gravity. In differential geometry, the Paneitz  
equation and its higher order $n$ generalizations 
have applications in the conformal geometry of $n$-manifolds, 
but no physical applications yet for general $n$. Interestingly, 
though, all the Paneitz equations have an interpretation
in terms of statistical mechanics.
\end{abstract}



\vfill
\smallskip
{\hrule\medskip
\noindent
{\copyright} (1999) The author. 
Reproduction for non-commercial purposes of this pre-print, 
in its entirety and by any means, is permitted. 
}

\eject

\noindent
\section{ INTRODUCTION}
\smallskip

Early in his scientific career Joel Lebowitz spent a postdoctoral year 
or so working with Lars Onsager. There are many scientific similarities
between these two towering figures of statistical physics: 
both in possession of penetrating mathematical powers; both 
broadly interested in physics; both with a good taste for choosing 
problems worthy to work on; and both with deep insights into 
physics that have inspired -- and continue to inspire -- many other
physicists and mathematicians around the world. Of course, Joel has 
also worked on many problems that were inspired by some work of
Onsager, and while it would be impossible for me here even to merely
mention them all, one contribution to a problem of Onsager that I was 
privileged to collaborate on with Joel has recently had some 
interesting mathematical spin-off, and I am very glad 
to present these new results in this paper in honor of 
Joel's  $70^{th}$ birthday. 

In the next section, I will first describe Onsager's application of
statistical mechanics to point vortices in two dimensions, the
questions it has raised, and our contribution to it. In the two
subsequent sections I will describe Nirenberg's problem in the
conformal differential geometry of two-dimensional manifolds
and its recent generalization to $n$-manifolds. Point
vortices in two dimensional flows have logarithmic pair 
interactions. In the last section I will explain how statistical 
mechanics of logarithmic interactions, but now in all dimensions,
provides solutions to the Nirenberg-type problems in conformal geometry. 

\medskip
\noindent
\section{ ONSAGER'S POINT VORTEX DISTRIBUTIONS}
\smallskip

In a pioneering paper on statistical fluid dynamics \cite{Onsager},
Onsager studied, among other things, the microcanonical ensemble 
of $N$ classical point vortices in a two-dimensional, 
incompressible Euler flow. The motion of $N$ vortices in a connected 
domain $\Lambda\subset\bbR^2$ is governed by Kirchhoff's Hamiltonian
\cite{KirchhoffBOOK,LinBOOK,MarchioroPulvirentiBOOKb},  
\beq
  H^{(N)}({\bf x}_1,...,{\bf x}_N)
 = {1\over 2} \sum~\sum_{\hskip-.7truecm 1\leq i\, ,\, j \leq N} 
c_i c_j G^*({\bf x}_i , {\bf x}_j )
+ \sum_{1\leq i \leq N} c_i F^{(N)}({\bf x}_i)\, ,
\label{hamiltonian}
\eeq
where ${\bf x}_i \in \Lambda$ is the position of the $i$-th vortex;
$c_i$ its circulation, expressed as dimensionless multiple 
of a suitable reference unit;   
$G^*({\bf x} , {\bf y} ) = G^*({\bf y} , {\bf x} )$ is a
renormalized Green's function for $-\Delta$ on $\Lambda$, i.e.
$G^*({\bf x} , {\bf y} ) = G({\bf x}, {\bf y})$ if 
${\bf x}\neq {\bf y}$, with $G({\bf x}, {\bf y})$ solving 
$-\Delta  G({\bf x}, {\bf y})= 2\pi \delta_{\bf y} ({\bf x})$ in $\Lambda$,
and 
\beq
G^*({\bf x},{\bf x}) = \lim_{{\bf y}\to {\bf x}} 
 \Big( G({\bf x},{\bf y}) + 
\ln \abs{{\bf x} - {\bf y} }\Bigr)\, .
\label{imagehamilton}
\eeq
In case of a bounded domain 
with piecewise regular boundary $\partial\Lambda$, 
the physically appropriate $0$-Dirichlet boundary conditions for $G$ 
are imposed on $\partial\Lambda$, but in principle other boundary conditions
are of interest too. Finally, $F^{(N)}$ is some externally applied 
stream function (whose strength may be proportional to $N$). 
Considering a bounded domain with finite area 
$|\Lambda|$, and noting that  up to a trivial factor $\sqrt{|c_i|}$
the canonically conjugate variables of the vortices are given 
by the Cartesian components 
of their positions in $\Lambda$,  Onsager observed that the 
phase space volume of the set $\{H^{(N)}<E\}$
in the $N$ vortices phase space $\Lambda^{N}$, 
\beq
\Phi_\Lambda(E,N) = \left| \left\{H^{(N)}<E\right\}\right| \, ,
\label{phasevolE}
\eeq
is a monotonically increasing function of $E$, bounded by $|\Lambda|^N$. 
He  noticed that this implies that Boltzmann's entropy
$S_\Lambda(E,N) = \ln\left(|\Lambda|^{-N}\Phi^\pr_\Lambda(E,N)\right)$
(where $\Phi^\pr_\Lambda(E,N) = \partial_E\Phi_\Lambda(E,N)$)
must reach a  maximum at a particular value $E_m(N;\Lambda)$ of energy, 
such that \cite{Onsager}, p. 281: ``negative ``temperatures''...will occur 
if $E>E_m$, ... vortices of the same sign will tend to cluster 
...[and] large compound vortices [be] formed in this manner...''

Onsager's insight not only predated Ramsey's prediction of negative
temperatures in spin systems \cite{Ramsey} by 7 years, Onsager was (once
again) way ahead of his time. His program was not picked up until 
almost a quarter century later, when Montgomery  pointed 
out \cite{MontgomeryPLA} that numerical simulations of turbulent fluid 
flows suggested an explanation in terms of Onsager's 1949 paper. 
But Onsager had not made any attempt to extract continuum 
vorticities from his statistical vortex distributions, 
and so several authors 
\cite{JoyceMontgomery,MontgomeryJoyce,PointinLundgren,LundgrenPointin,Kida}
now came up with the proposal to use mean-field theory for this purpose. 

As it is with traditions, and statistical mechanics surely 
has a long tradition, certain general wisdoms tend to be passed on 
which sometimes may not be so generally valid. One such general wisdom 
holds that `mean-field theory is wrong.' This traces back to the 
failure of the prototype mean-field theories of van der Waals
and Curie-Weiss to give the correct data for the critical 
point of the condensation and magnetization phase transitions, respectively.
In particular, they predict incorrect critical exponents. However,
as I recall Michael Fisher pointing out in a lecture on Coulombic
criticality, mean-field theory is not that bad after all. Another general 
wisdom states that thermodynamic concepts such as temperature become
precise only in the thermodynamic (bulk) limit of an infinitely
big system. A third general wisdom says that the statistical ensembles 
are equivalent. Against this background it is easier to appreciate 
that it took a while to realize that neither wisdom is correct 
in the case of Onsager's statistical theory of vortex clustering.

Indeed, the first attempt \cite{MontgomeryJoyce} to put Onsager's theory 
on a more rigorous basis was made in terms of the traditional 
bulk thermodynamic limit for a neutral two-species vortex 
system, using formal central limit arguments of Khinchin 
which where asserted to apply also to the negative temperature
regime. Moreover, in that paper  \cite{MontgomeryJoyce} also the BBGKY 
hierarchy was considered, and the mean-field equation for the 
distributions at negative $T$, which in \cite{JoyceMontgomery} had been 
obtained by the standard van-der-Waals-theory type combinatorics, was 
now obtained as Vlasov approximation. A subsequent rigorous study 
by Fr\"ohlich and Ruelle \cite{FroehlichRuelle} however showed that 
negative temperatures do, in fact, {\it not} exist in the
bulk thermodynamic limit of a neutral two-species vortex system. 

Inspection of the phenomenon that Onsager predicted reveals, however,
that $O(N_k^2)$ truly long range pair interactions in the $k^{th}$ 
cluster of vortices of the same sign are involved  \cite{LundgrenPointin}. 
Hence, we are dealing with a strictly nonuniform system with non-extensive 
energy scaling. Understood from this perspective, the
relevant limit $N\to\infty$ in which Onsager's prediction of negative 
vortex temperatures attains a rigorous meaning is not the bulk 
thermodynamic limit but rather a continuum (Euler) fluid
limit. Moreover, in this limit mean-field theory should become 
{\it exact} in the sense of a weak law of large numbers.  

Thus, the following picture emerges.
For the neutral two-species system, with $|c_i| = 1$, and
with $F\equiv 0$ for simplicity, distributed in 
a bounded domain $\Lambda\subset\bbR^2$ according to 
the microcanonical measure 
\beq
\mu^{(N,E)}(dx_1...dx_N) = {1\over \Phi^\pr_\Lambda(E,N)}
  \delta\left(E-H^{(N)}\right)dx_1...dx_N\, ,
\label{MCmeasure}
\eeq
where $dx$ denotes Lebesgue measure on $\bbR^2$,
we have to fix $\Lambda\subset\bbR^2$ and $\vareps = E/N^2 >0$. 
Then, in the limit $N\to\infty$, the Boltzmann entropy per 
vortex will converge to a continuous function of $\vareps$,
\beq
\lim_{N\to\infty} {1\over N} 
	\ln \Bigl(|\Lambda|^{-N} \Phi^\pr_\Lambda(N^2\vareps,N)\Bigr)
= s_{_\Lambda}(\vareps) \, ,
\label{MCentroplimguess}
\eeq
which is given by a variational principle, 
\beq
s_{_\Lambda}(\vareps)
=
\max_{\rho^+_{\phantom{+}}\! ,\, \rho^-_{\phantom{-}}}
 \sum_{\sigma = \pm} 
-\int_\Lambda \rho^{\sigma}\ln \big(|\Lambda|\rho^{\sigma}\big)	dx\, ,
\label{MCentroplimVPplumi}
\eeq
where the maximization is carried out over the  
probability densities $\rho^\pm_{\phantom{-}}$ 
which satisfy the energy constraint 
\beq
{1\over 2}\int_\Lambda\! \int_\Lambda
G({\bf x, y})\, \omega({\bf x})\,\omega({\bf y})dxdy = \vareps\, ,
\label{energyFUNCTL}
\eeq
where $\omega = \rho^+_{\phantom{-}} - \rho^-_{\phantom{-}}$.  
Moreover, the microcanonical equilibrium measure itself will
converge, for $\vareps >0$ and in a suitable topology, to a 
convex linear superposition of infinite products of those
absolutely continuous one-vortex measures whose densities $\rho_\vareps^\pm$
are the maximizers for
\refeq{MCentroplimVPplumi}-\refeq{energyFUNCTL}. In particular, 
if the superposition measure is a singleton (a unique maximizing pair
$\rho^\pm_\vareps$ exists), then we have the weak law of large numbers,
that for all $f\in C^{0}(\Lambda)$, 
\beq
\lim _{ N \to \infty}
 {2 \over N} \sum_{j=1}^{N/2}  f({\bf x}_{j}^\pm) 
= \int_\Lambda  f({\bf x}) \rho_{\vareps}^\pm({\bf x}) dx
\label{wlln}
\eeq
in probability. In \refeq{wlln}, the summation
extends over either all positive or all negative vortices. 
If the superposition is not a singleton, $\rho_\vareps^\pm$ in \refeq{wlln} 
is to be replaced by the corresponding  superposition of probabilities.
Furthermore, the maximizers satisfy the self-consistency conditions
\beq
\rho^\pm_\vareps({\bf x}) =  \exp\left(\beta\left[\mu_{ch}^\pm \mp
{ \int_\Lambda} G({\bf x, y})\omega_\vareps({\bf y })dy\right] \right)\, ,
\label{limitEQ}
\eeq
where $\omega_\vareps = \rho^+_\vareps - \rho^-_\vareps$, and where 
the constants $\beta$ and $\mu_{ch}^\pm$ are to be chosen so that 
the constraint $\int_\Lambda \rho^\pm_{\phantom{-}} dx = 1$ 
and \refeq{energyFUNCTL} are satisfied. 
Not only are the equations \refeq{limitEQ} clearly mean-field,
solutions having $\beta <0$ are known to exist, 
and the corresponding continuum vorticities $\omega_\vareps$ 
satisfy the stationary Euler equations for incompressible flows
\cite{MarchioroPulvirentiBOOKb}. 
Onsager's prediction of negative temperatures in 
vortex systems indeed attains a precise meaning in this limit.

There can be hardly any doubt that the above picture is correct,
and that it will eventually be verified rigorously. An almost complete 
verification is available for the mildly simpler single species system. 
By generalizing a result of Messer and Spohn \cite{MesserSpohn} for 
Lip\-schitz continuous interactions to the logarithmically singular
point vortex interactions, first the canonical point vortex 
ensemble was conquered, independently in \cite{clmpCMPcan}
and \cite{KiesslingCPAM}. While the microcanonical ensemble in 
the strict sense described above has not yet been treated, 
rigorous works exist in which the microcanonical point vortex measure 
$\delta(H-E)$ is replaced by a regularized measure, the 
regularization being removed after the limit $N\to\infty$ has been taken. 
Again, the first result, by Eyink and Spohn \cite{EyinkSpohn}, was for 
regularized interactions. The singular point vortex interactions where
then treated by Caglioti et al. \cite{clmpCMPmic}. The limit $N\to\infty$ 
in \cite{clmpCMPmic} is, however, constructed under the assumption 
that the microcanonical and canonical ensembles are equivalent, 
an assumption which is 
{\it not} generally valid for these finite-domain mean-field limits.  
This last barrier was finally overcome by Joel and myself 
in \cite{KieLebLMP}, where we go beyond equivalence of 
ensembles for the logarithmically singular \refeq{hamiltonian}. Of course,
the mean-field theory obtained is precisely
the single-species version of the one described above. 

Future work should extend these results to the neutral two-species 
vortex system. In contrast to the many rigorous results obtained for 
these neutral systems in the traditional thermodynamic limit,
see \cite{FroehlichRuelle} and references therein, only few 
facts are rigorously known for their Euler fluid limit. Curiously 
enough, while Fr\"ohlich and Ruelle \cite{FroehlichRuelle} proved absence 
of negative temperatures in the standard bulk limit, in 
\cite{KiesslingLMP} I was able to prove absence of positive temperatures 
in the Euler fluid limit for the neutral two-species system; 
in fact, $\beta <0$ is bounded away from zero. 
This raises the question as to the whereabouts of Onsager's $E_m$, 
at which temperatures are supposed to switch from positive to negative
values in a neutral system? The answer is that there is much room between
the low-energy regime $E=Ne$ of \cite{FroehlichRuelle} and the
high-energy regime $E=N^2\vareps$ of \cite{KieLebLMP,clmpCMPmic,EyinkSpohn}.
In fact, in \cite{campbelloneil} it was found that 
for a neutral two-species system $E_m(N) \sim C N\ln N$, 
falling precisely inbetween these two regimes.
The vicinity of $E_m$, analyzed in \cite{oneilredner,campbelloneil}, 
turns out to be a small-entropy regime where $S$, not $S/N$, converges
to a limit when  $N\to\infty$. 

The results in \cite{oneilredner,campbelloneil} are the only ones
obtained directly from $\delta(H-E)$. The construction of the 
Euler fluid limit $N\to\infty$ directly from $\delta(H-E)$ remains
an open problem. For now, \cite{KieLebLMP} is the final word in the 
construction of the Euler fluid limit for Onsager's statistical 
vortex distributions. 

Interestingly, \cite{KieLebLMP} also holds the key to some 
answers to questions in conformal geometry, to which we
turn next.

\bigskip
\noindent
\section{ NIRENBERG'S PROBLEM}
\smallskip

Many years ago, see \cite{MoserB}, 
Louis Nirenberg raised the following question:
``Which real functions $K(x)$ on $\bbS^2$ are Gauss curvature functions
for a surface over $\bbS^2$ whose metric $g$ is pointwise conformal to the 
standard metric $g_0$ on $\bbS^2$?'' To see what this question has
to do with Onsager's vortex distributions, we need to recall a few
basic concepts of differential geometry as found, for instance, in 
\cite{BandleBOOK,StruikBOOK}. As we shall see, the connection with
statistical mechanics, or at least thermodynamics, must have
been suspected by differential geometers long ago!

Consider first, for simplicity, surfaces over $\bbR^2$ that are embedded 
in $\bbR^3$. Thus, let ${\bf x}\in\Lambda\subset\bbR^2$. The
two Cartesian coordinates $x^1,x^2$ of ${\bf x}$
provide two independent real parameters for the local 
represention of a surface ${\cal S}\subset \bbR^3$, given by  
${\cal S} = X({\bf x})$, ${\bf x}\in\Lambda$, 
with $X\in \bbR^3$. 
The line element  on ${\cal S}$ is given by
$d\sigma^2 = g_{ij}dx^idx^j$, where we use Einstein's summation
convention, and where 
$g_{ij}= \langle \partial_{x^i}X,\partial_{x^j}X\rangle$
(Euclidean inner product) are the components of the metric 
tensor. If $\nu$ is the 
unit normal at ${\cal S}$ induced by the representation $X({\bf x})$,
then the Gauss curvature $K({\bf x})$ is defined by ($\times$ means
cross product in $\bbR^3$)
\beq
\partial_{x^1}\nu\times\partial_{x^2}\nu =
K({\bf x})\, \partial_{x^1}X\times\partial_{x^2}X
\label{gaussCURV}
\eeq
Gauss' {\it Theorema Egregium} asserts that $K({\bf x})$ depends only
on the $g_{ij}$. Writing $g_{11}=E$, $g_{12}=F$, and $g_{22}=G$,
and moreover $x^1=s$ and $x^2=t$, we have the Frobenius formula
\bea\label{FrobeniusK}
K =&-& {1\over 4(EG-F^2)^2}\, {\rm det}
\pmatrix{E&F&G\cr E_s&F_s&G_s\cr E_t&F_t&G_t\cr} \nonumber\\
&+&  {1\over 2\sqrt{EG-F^2}} \left[ 
\partial_s\left( {F_t-G_s\over \sqrt{EG-F^2}}\right) 
-
\partial_t\left( {E_t-F_s\over \sqrt{EG-F^2}}\right) 
				\right]
\eea
see \cite{BandleBOOK,StruikBOOK},
from which it follows that $K$ is independent of 
the parametric representation $X({\bf x})$ of ${\cal S}$. This 
freedom can be used to simplify \refeq{FrobeniusK}. In particular, 
the representation for which  $E=G$ and $F=0$ is conformal, 
i.e. $d\sigma^2 = Ed{\bf x}^2$. Interestingly, it is  known 
in differential geometry as {\it isothermic} parameter representation. 
In this representation, \refeq{FrobeniusK} reduces to
$K = - {1\over 2E} \left( \partial^2_s \ln E +  \partial^2_t \ln E \right)$,
or, setting $\ln E = 2u$ and recalling $(s,t)={\bf x}$,
\beq
K({\bf x}) = -e^{-2u({\bf x})}\Delta u({\bf x})  
\label{isothermicPDE}
\eeq
which gives us $K$ when $u$ (i.e. $E$) is given. Nirenberg's 
question, here its analog on $\bbR^2$, addresses the inverse problem, i.e. 
to prescribe the putative Gauss curvature function $K$ and study 
\refeq{isothermicPDE} as a nonlinear elliptic PDE for the unknown
$u({\bf x})$, ${\bf x}\in\Lambda\subset\bbR^2$. 
In particular, when $K$ is constant, $K=\pm 1$ by scaling, 
then \refeq{isothermicPDE} is known 
in differential geometry as Liouville's equation \cite{Liouville}.
Clearly, Liouville's equation is  identical to what in statistical 
physics goes under the name Poisson-Boltzmann equation,
\beq
-\Delta \psi({\bf x}) = 2\pi e^{\beta[\mu_{ch}-\psi({\bf x})]}
\label{PBeq}
\eeq
for the spatial density of a `perfect gas' in $\Lambda\subset\bbR^2$, 
in thermal equilibrium at temperature $\beta^{-1}$ 
and chemical potential $\mu_{ch}$, distributed in its own Coulomb 
(or Newton) potential field $\psi$. Hence, the notion of 
`isothermic parameters,' so it seems; however, I have not been able 
to trace the originator of this terminology. In any event, \refeq{PBeq} is 
of course identical to the one-species specialization of \refeq{limitEQ}, 
i.e. \refeq{limitEQ} with $\rho^+\equiv \rho$ and $\rho^-\equiv 0$, and with 
$\psi({\bf x})= \int_\Lambda G({\bf x, y})\rho({\bf y })dy$ 
the stream function, so that $-\Delta \psi({\bf x})= 2\pi \rho({\bf x })$.  
We have come back full circle to Onsager's problem.

Liouville \cite{Liouville} himself already showed that the equation derived 
by him is in a certain sense completely integrable. In particular, 
all entire solutions on $\bbR^2$ with finite integral curvature
$\kappa = \int_{\bbR^2} K e^{2u({\bf x})}dx$ have been identified;
namely, no entire solution exists when $K=-1$ \cite{Poincare,Osserman},
while for $K=+1$ Liouville's equation is solved when $\exp(2u)$ 
is any point on the conformal orbit of the Jacobian of the 
stereographic map $\bbS^2\to\bbR^2$, and $\kappa = 4\pi$  for all
solutions, then; see \cite{ChenLiA,CarlenLossPAPb,ChouWan,ChaKieGAFA}. 

For the prescribed Gauss curvature problem in all $\bbR^2$ 
with $K\neq const.$, no general solution of \refeq{isothermicPDE} 
is available, yet  over the years a vast knowledge has accumulated, 
e.g. \cite{Sattinger}, \cite{Ni}, \cite{McOwen}, \cite{Aviles}, 
\cite{ChengNi}, \cite{ChenLiB}, \cite{ChaKieCMP}, \cite{ChengLin},
\cite{ChaKieDMJ}, \cite{PrajapatTarantello}.  
Of particular interest are radially symmetric Gauss curvature 
functions, for then the interesting question arises whether 
solutions $u$ exist that break the radial symmetry.  In the 
following, the sign of $K$ (as a function) is defined as: 
$\sign (K) =1$ if $K\not\equiv 0$ 
and $K({\bf x})\geq 0$ for all ${\bf x}$;  $\sign (K) = -1$
if $K\not\equiv 0$ and $K({\bf x})\leq 0$ for all ${\bf x}$;
and $\sign (K) = 0$ if $K\equiv 0$. In all other cases $\sign(K)$
is not defined. The next theorem is taken from \cite{ChaKieDMJ}.

\smallskip
\noindent
\proclaim{Theorem}\thmlbl\ChaKieTHM:\ {\it 
	Assume $K\in C^{\alpha}(\bbR^2)$ is radially symmetric, has
	well-defined sign, and satisfies
\beq
\int_{\bbR^2} |K({\bf x})| e^{2h({\bf x})}|{\bf x}|^{\lambda} dx < \infty\, 
\label{KconditionA}
\eeq
	for some non-constant, harmonic function 
	$h:\bbR^2\mapsto \bbR$, and some $\lambda >0$. Assume also that
\beq
\int_{{\rm B}_1(y)}
|y - x|^{-\gamma} |K(x)|e^{2H(x)} dx \longrightarrow 0
\qquad	{\rm as}\ |y| \to \infty\, ,
\label{KconditionB}
\eeq
	for all $0 < \gamma < 2$. If $\sign(K)=-1$, define
\beq
\kappa^*(K;h) = -2\pi 
\sup \bigl\{\lambda>0 :\ {\refeq{KconditionA}\ is\ satisfied\, } \bigr\}
 \label{extremekappa}
\eeq
	where $\kappa^*$ might be $-\infty$ for some $K\leq 0$.  
	Then, for any such $K$ and $h$, and $\kappa$ satisfying 
\beq
\kappa \in \cases{
\ (\kappa^*,0)		&{\rm if}\ $\sign(K)= -1$;\cr
\qquad     [0]		&{\rm if}\ $\sign(K)=\ \  0$;\cr
\qquad     (0,4\pi)  	&{\rm if}\ $\sign(K)= +1$;\cr}
\label{kappaRANGE}
\eeq
 there exists a classical solution $u = U_{h,\kappa}$ of
\refeq{isothermicPDE} with prescribed Gaussian curvature function $K$, 
uniquely if $K \leq 0$, having integral curvature
\beq
\int_{\bbR^2} K({\bf x}) e^{2U_{h,\kappa}({\bf x})}dx \, ={\kappa}\, 
\label{kappaINTcurv}
\eeq
and asymptotic behavior 
\beq
\qquad
U_{h,\kappa}(x) 
= h({\bf x}) - {\kappa \over 2\pi} \ln |{\bf x}| + o(|\ln |{\bf x}||)  
\qquad {\rm as}\ |{\bf x}| \to \infty\, .
\label{asympFORMULAu}
\eeq
}
\smallskip
Theorem \ChaKieTHM\ considerably generalizes  an earlier result of 
this kind (Thm.III in \cite{ChengNi}) which is restricted to 
compactly supported $K$ in $\bbR^2$. 
Theorem \ChaKieTHM\ is proven in \cite{ChaKieDMJ} as corollary 
of the construction, also given in \cite{ChaKieDMJ}, of the Euler fluid limit 
in $\bbR^2$ for the canonical point vortex ensemble, which  generalizes
the one for finite domains $\Lambda\subset \bbR^2$ 
\cite{KiesslingCPAM,clmpCMPcan}. The canonical point vortex measure on 
$\bbR^{2N}$ is given by  
\beq
\mu^{(N,\beta)} (dx_1...dx_N) = {1\over Z(\beta,N)}
  \exp\left(-\beta N^{-1}  H^{(N)}\right)dx_1...dx_N\, ,
\label{Cmeasure}
\eeq
with $G({\bf x,y}) = -\ln |{\bf x -y}|$ and 
$F^{(N)}({\bf x})= -N\beta^{-1}[\ln |K({\bf x})| + 2h({\bf x})]$
in $H^{(N)}$, \refeq{hamiltonian}.  The reciprocal
temperature $\beta$ in \refeq{Cmeasure} and the integral curvature $\kappa$ 
in Theorem  \ChaKieTHM\ are related by $\kappa = -\beta \pi$, 
with $-\beta \in (\kappa^*/\pi,4)$.

Encouraged by these achievements of the canonical statistical mechanics
approach to the prescribed Gauss curvature problem on $\bbR^2$, 
we now return to Nirenberg's problem in its original setting on
the sphere $\bbS^2$. Writing the conformal deformation of the metric as 
$g= e^{2u(x)}g_0$, where $g_0$ is the standard metric on $\bbS^2$, the 
problem is to find all $K(x)$, $x\in \bbS^2$, for which
\beq
-\Delta u(x) = K(x)e^{2u(x)} - 1
\label{GaussKsphereEQ}
\eeq
has a solution $u$ on $\bbS^2$. Here, $\Delta$ is the Laplace-Beltrami
operator on $\bbS^2$ w.r.t. $g_0$. Superficially 
\refeq{GaussKsphereEQ} is hardly any different from the prescribed 
Gauss curvature equation on $\bbR^2$ \refeq{isothermicPDE}. However,  
Nirenberg's problem for $\bbS^2$ is a hard problem, indeed;
see \cite{MoserB}, \cite{KazdanWarner}, \cite{ChangYangA}, \cite{ChangYangB}, 
\cite{Han}; see also \cite{KazdanWarner}, \cite{DJLW} for more general 
compact Riemann surfaces. 
 
There are many obstructions to finding admissible $K$ for
\refeq{GaussKsphereEQ}.
The celebrated Kazdan-Warner theorem \cite{KazdanWarner} is one of them.
According to \cite{KazdanWarner} a Gauss curvature function  $K$ on $\bbS^2$
cannot be axially symmetic {\it and} monotonic, unless 
it is a constant function, say $K\equiv 1$. In the latter case  
the problem is completely solved, \cite{MoserA}, \cite{Aubin}, \cite{Onofri}.
Another, more serious obstruction is the famous Gauss-Bonnet theorem 
\cite{StruikBOOK}, which relates the Gauss curvature $K_g$ on a general 
compact $2$-manifold $(M^2,g)$ without boundary to the Euler-Poincar\'e 
characteristic $\chi(M^2)$, a topological invariant, by
\beq
\chi(M^2) = 
{1\over 2\pi} \int_{M^2} K_g d{\rm vol}_g \, .
\label{gaussbonnetzwei}
\eeq
Since 
$\chi(\bbS^2)>0$, it follows from \refeq{gaussbonnetzwei} that 
$K(x)$ has to be positive for some $x\in\bbS^2$, \cite{MoserB}. 
Furthermore, we have $\chi(\bbS^2)=2$. Thus, $\kappa = 4\pi$, i.e. 
\beq
 \int_{\bbS^2} K(x)e^{2u(x)}dx = 4\pi\, , 
\label{intCURVsphere}
\eeq
which of course results also  by directly
integrating \refeq{GaussKsphereEQ} over $\bbS^2$ 
Contrast \refeq{intCURVsphere} with the range of integral curvatures 
covered in Theorem \ChaKieTHM! In particular, notice that 
$\kappa = 4\pi$ is {\it not} included in Theorem \ChaKieTHM,
and this is a matter of principle. Indeed, the restriction to
$-\beta < 4$ in the canonical ensemble is an obstruction 
imposed on us by the {\it local} integrability requirement 
of the singularities in \refeq{Cmeasure}, and this will not improve
if we put our vortex system on the sphere \cite{KieSpohnCMP}. Thus, 
\refeq{intCURVsphere} dashes any hope to apply the Euler fluid limit 
of the canonical point vortex ensemble at $\beta =-4$ 
to the prescribed Gauss curvature problem on $\bbS^2$. 

Not all hopes are dashed, though. Physically speaking, what
happens at $-\beta =4$ in the finite $N$, single-species
canonical ensemble is that the  vortex system concentrates
onto a single point. This is done at the expenses of the 
`heat bath,' which delivers a positive infinite amount of 
energy into the vortex system at fixed negative temperature. 
Taking the limit $N\to\infty$ of this  concentrated singular state gives a
Dirac $\delta$ mass on $\bbS^2$, corresponding to $\vareps =+\infty$.
However, performing the mean-field limit $N\to\infty$ first for
$-\beta < 4$ produces a regular solution of the resulting analog 
on $\bbS^2$ of the Poisson-Boltzmann equation \refeq{PBeq}, and subsequently 
letting $-\beta \nearrow 4$ may, or may not result in a regular 
limiting solution at $-\beta =4$. This can be made more precise with
the so-called concentration-compactness alternative of P.L. Lions, 
compactness of the sequence of solutions as  $-\beta \nearrow 4$ 
implying a regular limiting solution at $-\beta =4$. The canonical
ensemble may nevertheless not provide enough control to 
decide the concentration-compactness alternative. However, 
whenever compactness holds, a regular solution exists which 
has finite energy $\vareps$, and some finite entropy $s$. 
%
%
We conclude that the study of the mean-field limit
for the microcanonical ensemble of point vortices on $\bbS^2$,
with external stream function $F^{(N)}$, can be expected to
yield valuable new information on Nirenberg's problem. 

As remarked earlier,  the microcanonical ensemble in its strict 
sense has not yet succumbed to rigorous analysis. However, 
for the  prescribed Gauss curvature problem on $\bbS^2$ the
construction given in \cite{KieLebLMP} is fully sufficient. As a matter
of fact, we shall state our microcanonical results in the more
general setting on $\bbS^n$, $n\geq 2$. To see what this now is
about, we need to briefly digress into Paneitz theory.

\medskip
\noindent
\section{ PANEITZ EQUATIONS}
\smallskip

Initiated by the conformal covariance in Minkowski space-time 
$\bbR^{1,3}$ of the Maxwell equations of electromagnetism,  
S. Paneitz \cite{Paneitz} in 1983 discovered a quartic conformally 
covariant differential operator for arbitrary pseudo-Riemannian 
$1,3$-manifolds, together with an associated new conformal curvature 
invariant. Recently this theory has received considerable attention,
see \cite{BransonA}, \cite{BCY}, \cite{ChangYangC}, \cite{BransonB}, 
\cite{Chang}, \cite{ChangQing}.
The significance of the  Paneitz curvature,  $Q_g$,  becomes apparent 
through an analog of the Gauss-Bonnet formula for a
general compact $4$-manifold $(M^4,g)$ without boundary, \cite{ChangYangD}, 
\beq
\chi(M^4) = 
{1\over 8\pi^2} \int_{M^4}\left( {1\over 4} |W(x)|^2 +Q_g(x)\right)
d{\rm vol}_g
\label{gaussBonnetfour}
\eeq
Here, $\chi(M^4)$ is the Euler-Poincar\'e characteristic of $M^4$, 
and $W$ is the pointwise conformally invariant Weyl tensor. 
Moreover, $Q_g$ is defined  by
\beq
6 Q_g(x) = -\Delta_g R_g(x) + {1\over 4} R_g^2(x) - 
3 |\widetilde{\rm Ric}_g(x)|^2
\label{QonMdef}
\eeq
where $\Delta_g$ is the Laplace-Beltrami operator on $(M^4,g)$, 
$R_g$ is the scalar curvature and $\widetilde{\rm Ric}_g$ the 
traceless Ricci tensor,  see \cite{Paneitz}, \cite{Chang}.
Associated with $Q_g$ is Paneitz' quartic conformally covariant
operator, such that, given $(M^4,g_0)$ and a conformal change
of metric written as $g= e^{2u(x)}g_0$, $x\in (M^4,g_0)$, the new
curvature $Q_g$ is given by
\beq
 Q_g(x) = e^{-4u(x)}  
\left( \Delta^2_{g_0} u(x) +\delta_{g_0}\left( {2\over 3} R_{g_0}(x)I 
- 2 {\rm Ric}_{g_0}(x)\right)d_{g_0} u(x) + Q_{g_0}(x)\right)
\label{PaneitzEQ}
\eeq
Here, $d$ is the differential and $\delta$ the divergence. 

An analog of Nirenberg's problem on the $4$-manifold $(\bbS^4,g_0)$, 
with $g_0$ the standard metric, can be formulated thus: 
``Which real functions $Q(x)$, $x\in\bbS^4$, are Paneitz curvature 
functions for a $4$-manifold whose metric is pointwise conformal to
the standard one?'' This analog of Nirenberg's problem can be rephrased 
in terms of \refeq{PaneitzEQ}, namely to find all functions $Q(x)$ on $\bbS^4$ 
such that the fourth-order PDE 
\beq
\Delta^2 u(x) -2\Delta u(x) = Q(x) e^{4u(x)} - 6
\label{PanEQfour}
\eeq
has  a solution $u$ on $\bbS^4$. 
Generalizations to $\bbS^n$ (and, hence, to $\bbR^n$) 
of the Paneitz equation \refeq{PanEQfour} have been derived as well. 
On $\bbS^n$, $n\geq 2$,  we have
\beq
P_n u(x) = Q(x) e^{nu(x)} - (n-1)!
\label{PanEQn}
\eeq
with 
\beq
P_n = \cases{ \displaystyle
\phantom{\sqrt{-\Delta + \left({\textstyle n-1\over 2}\right)^2}\quad }
\prod_{k=0}^{n-2\over 2}(-\Delta +k(n-k-1)) ;\quad n\
{\rm even}\cr\cr
\displaystyle
\sqrt{-\Delta + \left({\textstyle n-1\over 2}\right)^2}\quad 
\prod_{k=0}^{n-3\over 2}(-\Delta +k(n-k-1)) ;\quad n\
{\rm odd},\cr\cr}
\label{PanOPn}
\eeq
see \cite{CarlenLossPAPb}, \cite{Beckner}, \cite{ChangYangE}, \cite{ChangQing}.
On $\bbR^n$, we simply have 
\beq
(-\Delta)^{n/2} u(x) = Q(x) e^{nu(x)}
\label{PanEQnflat}
\eeq
with $x\in \bbR^n$ now.
The operator $P_n$ in \refeq{PanEQn} is the Paneitzian, and
$Q(x)$ the Queervature (pardon the pun) of order $n$.
Notice that for $n$ odd, $P_n$ is a pseudo differential operator. 

\bigskip
\noindent
\section{LOGARITHMIC INTERACTIONS IN ALL DIMENSIONS}
\smallskip

The increased complexity of the operators $P_n$ given in \refeq{PanOPn} 
for high dimensions gives the Paneitz equations \refeq{PanEQn} 
a formidable appearance.  Also \refeq{PanEQnflat} is not too
familiar when $n >2$. However, notice that 
the resolvent kernel of $P_n$ on $\bbS^n$ ($\bbR^n$), with $P_n$
restricted to the orthogonal complement of its kernel space,
is always $G(x,y) = -\ln|x-y|$, with $x,y\in\bbS^n$
$(x,y\in\bbR^n)$, and  with $|\ .\ |$ the chordal 
distance on $\bbS^n$ (Euclidean distance in $\bbR^n$), cf.\cite{CarlenLossPAPb}. 
More precisely, 
$-P_n\ln|x-y| = (1/2)(n-1)!|\bbS^n|(\delta_y(x) - |\bbS^n|^{-1})$
on $\bbS^n$.
Hence, all the equations \refeq{PanEQn}, as well as \refeq{PanEQnflat},  
have a statistical mechanics interpretation whenever $Q$ has a 
well defined sign.

In the following we discuss the prescribed Paneitz curvature problems
on $\bbS^n$, \refeq{PanEQn}, \refeq{PanOPn}, using the mean-field limit of
the regularized microcanonical ensemble of \cite{KieLebLMP}. 
Nirenberg's problem on $\bbS^2$ is contained in the analysis 
as special case $n=2$. We also remark on the equations in $\bbR^n$, 
\refeq{PanEQnflat}, using just the canonical ensemble.  

Let us begin with the simpler \refeq{PanEQnflat}. A brief moment of
reflection reveals that Theorem \ChaKieTHM, {\it and its proof}, 
have an immediate generalization to the Paneitz equation 
\refeq{PanEQnflat}. In \refeq{hamiltonian} we then have to set 
$G(x,y)= -\ln|x-y|$, $x,y\in \bbR^n$.
We also replace $|K|$ in $F^{(N)}$ by $|Q|$, and
$2h$ by $nh$, where instead of a harmonic function as
in Theorem \ChaKieTHM\ now $h$ is a non-constant element 
of the kernel space of $(-\Delta)^{n/2}$ on $\bbR^n$, to which 
we may want to refer as higher harmonic function. Moreover,
$Q$ shall have well defined sign and satisfy  analogous 
integrability conditions as \refeq{KconditionA} and
\refeq{KconditionB} in Theorem \ChaKieTHM. 
Finally, the critical $\beta = -4$ for the canonical ensemble
changes to $\beta = - 2n$, and the corresponding critical 
integral Gauss curvature $\kappa = 4\pi$ in  Theorem \ChaKieTHM\ 
changes to a critical integral Paneitz curvature 
$q = (n-1)! |\bbS^{n}|$, where $q= \int_{\bbR^n} Q(x) e^{n u(x)}dx$. 
These numbers are also the sharp 
constants in the Trudinger-Moser type inequalities on $\Lambda\subset\bbR^n$ 
and $\bbS^n$, \cite{Trudinger,MoserA,Beckner,CarlenLossPAPb}. 
  
We now come to the problem on $\bbS^n$. Since $\bbS^n$ is a 
compact manifold without boundary we do not need any delicate 
unbounded-domain estimates of the sort needed in \cite{ChaKieDMJ} or 
\cite{KieSpohnCMP}. The technique of \cite{KieLebLMP} carries
over to $\bbS^n$ without major changes. In \refeq{hamiltonian} 
we simply have to set $G(x,y)= -\ln|x-y|$, $x,y\in \bbS^n$.
We also let $F^{(N)}(x)=Nf(x)$, with $f$ some continuous 
function on $\bbS^n$ satisfying $\int_{\bbS^n}f(x)dx =0$. 
Notice that not all $f$ will correspond to a Paneitz curvature function.  

Our regularized microcanonical probability
measure on $(\bbS^n)^N$ is of the form \cite{KieLebLMP}
\beq
   \mu^{(N,\vareps,\sigma)}(dx_1...dx_N) = {1\over Z} \exp
\Bigl[- N{1\over 2\sigma^2}\Bigl(\vareps - {1\over N^2} H^{(N)}\Bigr)^2
\, \Bigr]   dx_1...dx_N\, ,
\label{gaussmeas}
\eeq
where $\sigma>0$ and $\vareps$ are fixed real numbers, and
\beq
{Z}(N,\vareps,\sigma) =  {\int_{(\bbS^n)^N} } \exp
\Bigl[- N{1\over 2\sigma^2}\Bigl(\vareps - {1\over N^2} H^{(N)}\Bigr)^2
\Bigr] dx_1...dx_N \, .
\label{gaussstrucfunc}
\eeq
The Hamiltonian is given in \refeq{hamiltonian}, with the identifications
of $G$ and $F$ mentioned above. The microcanonical ensemble at fixed
$N$ is obtained in the limit $\sigma \to 0$ in \refeq{gaussmeas}, 
giving a delta measure concentrated on $\{ H^{(N)} = E\}$, with 
$E = N^2 \vareps$, as can be easily verified using geometric measure 
theory. 

Let $P({(\bbS^n)^N})$ denote the probability measures on ${(\bbS^n)^N}$.
For any $N\in\bbN$, we define the entropy of $\varrho_N \in P({(\bbS^n)^N})$
relative to the normalized uniform measure $|\bbS^n|^{-N}dx_1...dx_N$ by
\beq
{\bf S}(\varrho_N) 
= -\int_{{(\bbS^n)^N}} \rho_N\ln\left(|\bbS^n|^N\rho_N\right) dx_1...dx_N \, ,
\label{entropyN}
\eeq
if $\varrho_N$ is absolutely continuous w.r.t. uniform
measure on ${(\bbS^n)^N}$,
having density $\rho_N$, and provided the integral on the r.h.s.
of \refeq{entropyN} exists; ${\bf S}(\varrho_N) = -\infty$
in all other cases.  For $\varrho_1=\varrho \in P(\bbS^n)$, 
we define a  one-particle penalized entropy functional by
\beq
{\bf R}_{\vareps,\sigma}(\varrho) =  {\bf S}(\varrho) 
 - {1\over 2\sigma^2} \left( \vareps - 
{1\over 2} \int_{\bbS^n} \int_{\bbS^n} 
G({ x,y})\varrho({d x})\varrho({d y})
- \int_{\bbS^n} f(x)\varrho(dx) \right)^2  
\label{penaltyentropy}
\eeq
for those $\varrho(dx) = \rho\, dx$, with $\rho$
a probability density, for which ${\bf S}(\varrho) > -\infty$. 
We set ${\bf R}_{\vareps,\sigma}(\varrho) = -\infty$ in all
other cases. Here,  ${\bf S}(\varrho) = {\bf S}(\varrho_1)$
is the one-particle entropy as defined in \refeq{entropyN}.
We write $M_{\vareps, \sigma}$ for the 
set $\{\varrho_{\vareps,\sigma}\}\subset P(\bbS^n)$ 
of maximizers of ${\bf R}_{\vareps,\sigma}$. 

By $\Omega=(\bbS^n)^\bbN$ we denote the $\bbS^n$-valued 
infinite exchangeable sequences, by $P^{sym}(\Omega)$ the 
permutation-invariant probability measures on $\Omega$.
According to the theorem of de Finetti  -- Dynkin \cite{Finetti,dynkin}
every $\mu\in P^{sym}(\Omega)$ is a unique
convex linear superposition of product measures, see also
\cite{Diaconis} and \cite{hewittsavage}. By a theorem of Hewitt and 
Savage \cite{hewittsavage}, the product states $\varrho^{\otimes \bbN}$ 
are also the extreme points of the convex set $P^{sym}(\Omega)$. 
Hence, we have the extremal decomposition 
\beq
  \mu_k(dx_1...dx_k) = \int_{P(\bbS^n)} \nu(\mu\vert d\varrho) 
\, \varrho^{\otimes k}(dx_1...dx_k)\, ,
\label{productdecomp}
\eeq
with $\mu_k(dx_1...dx_k)\in P({(\bbS^n)}^k)$ the $k$-th marginal 
measure of $\mu$.

As in \cite{KieLebLMP}, we first take the limit $N\to\infty$ of 
\refeq{gaussmeas} for fixed $\vareps$ and $\sigma$. This 
gives us mean-field theory with a two-parameter penalized 
entropy principle, as follows.

\smallskip
\proclaim{Theorem}\thmlbl\Gmeasurelimit:\ {\it
 For each $\vareps\in\bbR$ and $\sigma >0$ fixed, 
one has
\beq
\lim_{N\to\infty} {1\over N}\ln \left[|\bbS^n|^{-N}
Z(N,\vareps,\sigma) \right]
= {\bf R}_{\vareps,\sigma}(\varrho_{\vareps,\sigma}) 
\label{Gentroplimexist}
\eeq
with $\varrho_{\vareps,\sigma} \in M_{\vareps, \sigma}$. 
Moreover, \refeq{gaussmeas} has at least one limit point
in the corresponding subset of $P^{sym}(\Omega)$, 
convergence understood for all the marginals in the sense of Kolmogorov,
\cite{BillingsleyBOOK}, here weakly in $L^p$, $p<\infty$. 
The decomposition measure $\nu(\mu^{(\vareps,\sigma)}|d\varrho)$ 
of any limit point $\mu^{(\vareps,\sigma)}$ is
concentrated on $M_{\vareps,\sigma}$.}

The subsequent limit $\sigma\to 0$ now gives the anticipated 
mean-field variational principle for the microcanonical entropy.
We denote by $L^{1,+}_1(\bbS^n)$ the subset of the positive cone
of $L^1(\bbS^n)$ whose elements $\rho$ satisfy 
$\int_{\bbS^n} \rho\, dx =1$, and by 
$L^{1,+}_{1;\vareps}(\bbS^n)$ the subset of 
$L^{1,+}_1({\bbS^n})$ for which
\beq
{1\over 2} 
\int_{\bbS^n}\int_{\bbS^n} G(x,y)\varrho(dx)\varrho({d y})
+ 
\int_{\bbS^n}f(x)\varrho(dx) = \vareps\, ,
\label{fluidenergy}
\eeq
with $\varrho(dx) = \rho(x)dx$. 
Let $\vareps_0(f)$ denote the minimum over $P(\bbS^n)$ of the 
functional on the left side of \refeq{fluidenergy}.
\smallskip

\proclaim{Theorem}\thmlbl\MCsigmalimit:\ {\it
Fix $\vareps \in \bbR$.
 
Part 1. Let $\vareps \geq \vareps_0$. Then  the limit
\beq
s(\vareps) = \lim_{\sigma \to 0}
{\bf R}_{\vareps,\sigma}(\varrho_{\vareps,\sigma}) 
\label{sigmalimexist}
\eeq
exists and satisfies the variational principle
\beq
s(\vareps) = \max\, \left\{{\bf S}(\varrho)|\;
\varrho(dx) = \rho\, dx;\, \rho\in L^{1,+}_{1;\vareps}\right\}\, .
\label{MCmaxRHOvp}
\eeq
If $\vareps >\vareps_0$, all maximizers $\rho_\vareps$ for
\refeq{MCmaxRHOvp} satisfy the Euler-Lagrange equation
\beq
\rho_\vareps(x) = 
 \exp\left(\beta\left[\mu_{ch} -
\int_{\bbS^n} G(x,y)\rho_\vareps({y}) dy 
- f(x) \right] \right)\, ,
\label{MCeulerlagrange}
\eeq
where $\beta$ and $\mu_{ch}$ are real Lagrange parameters for the
constraints $\rho\in L^{1,+}_{1;\vareps}$. For $\vareps=\vareps_0$,
the maximizer(s) solve the free boundary problem 
obtained from \refeq{MCeulerlagrange} in the limit $\beta\to+\infty$.

Furthermore, let $M_{\vareps}$ denote the 
set of maximizers $\rho_\vareps$ for \refeq{MCmaxRHOvp}.
Let $\mu^{(\vareps)}$ be a weak limit point, as $\sigma\to 0$,
of the measure $\mu^{(\vareps,\sigma)}$. Then
$\mu^{(\vareps)}\in P^{sym}(\Omega)$, and its decomposition measure
$\nu(\mu^{(\vareps)}|d\varrho)$ is concentrated on $M_{\vareps}$. 

Part 2. Let $\vareps < \vareps_0$. In this case 
$\lim_{\sigma \to 0} 
{\bf R}_{\vareps,\sigma}(\varrho_{\vareps,\sigma}) 
= -\infty$. }
\smallskip

The proofs of Theorems \Gmeasurelimit\ and \MCsigmalimit\  are nearly
verbatim copies of the proofs for two-dimensional domains given in
\cite{KieLebLMP}, with a few functional analytical differences. 
Details will appear elsewhere.

Our \refeq{MCeulerlagrange} is the dual equation to \refeq{PanEQn}
 on $\bbS^n$. The differential geometric problem is to find such $f$ for which 
\refeq{MCeulerlagrange} has a solution with $\beta = - 2n$, in which case 
$(n-1)!|\bbS^{n}|\exp\left(2nf(x)\right)$
can be identified with a prescribed Paneitz curvature functions $Q(x)$. 
Clearly, only such $Q$ with $\sign(Q) =+1$ can be found this way. 
The parameter ${\beta\mu_{ch}}$, which accounts for the 
requirement that $\rho$ is a probability density, is simply absorbed in 
$u(x)$, but the parameter $\beta$ in \refeq{MCeulerlagrange} is implicitly
determined by the choice of $\vareps$. Hence, the problem becomes: 
``Find all $f$ for which the map $\vareps\mapsto\beta(\vareps)$ 
takes the value $-2n$.''

One obvious such $f$ is $f\equiv 0$, in which case the equation 
\refeq{MCeulerlagrange} with $\beta =-2n$ is completely 
integrable \cite{CarlenLossPAPb,Beckner}. Namely, \refeq{MCeulerlagrange} 
is then invariant under the full conformal group on $\bbS^n$, 
having a unique solution --- up to rotations and reflections on
$\bbS^n$ --- {\it for each} $\vareps \geq \vareps_0$.

The problem to find admissible $f\not\equiv 0$ 
may not appear any simpler than the original Nirenberg problem 
and its generalization to higher $n$, but now
thermodynamics comes to the rescue.
It is straightforward to show that $s(\vareps)$ is  continuous
and piecewise differentiable and that ${{\rm Ran}(\partial_\vareps s)}$
is connected. As in ordinary thermodynamics, so also here we have 
the identification 
\beq
\beta = \partial_\vareps s(\vareps)
\label{entropyderivative}
\eeq
wherever the derivative is defined.
We have $\partial_\vareps s(\vareps)\to +\infty$ as
$\vareps\to\vareps_0^+$, except when $f\equiv 0$. Moreover, 
for all $f\not\equiv 0$ we have
$s(\vareps_\infty ) = 0$ and $\partial_\vareps s(\vareps_\infty) = 0$, 
where $\vareps_\infty  = (1/2)\int_{\bbS^n}\int_{\bbS^n}
G(x,y)|\bbS^n|^{-2}dxdy$. For  $\vareps >\vareps_\infty$, we have 
$\partial_\vareps s(\vareps) <0$.  
Therefore, it suffices to solve the  simpler problem of finding those $f$ 
for which $\partial_\vareps s(\vareps) < -2n$ for some large enough 
$\vareps>\vareps_\infty$. Using physical intuition as guidance,
a little reflection reveals that a solution to the generalized
Nirenberg problem should exist whenever $f$ allows 
particles to cluster in at least two spatially separated regions 
when $\beta <0$, or, in technical language:

\proclaim{Conjecture}\thmlbl\curvaturef:\ {\it
Any $f\in C^0(\bbS^n)$ satisfying $\int_{\bbS^n}f(x) dx = 0$ which
has at least two isolated maxima can be identified with a Paneitz 
curvature function by $Q(x) = (n-1)!|\bbS^{n}|\exp\left(2nf(x)\right)$.}
\smallskip

A variant of Conjecture \curvaturef\ for $\bbS^2$ with 
somewhat stronger regularity assumptions on the Gauss curvature
has been proven with PDE techniques in \cite{ChangYangA}, see their 
Theorems I and II. Our microcanonical mean-field limit now offers 
a new perspective on proving a result like Conjecture \curvaturef\ ---
simultaneously in all dimensions. I hope to report on the details 
of such an effort in some future publication. 

As a final remark, I mention that also the ground state problem
$\vareps=\vareps_0$,
while only indirectly relevant to our conformal geometric problems 
on $\bbS^n$, is of quite some interest, in other contexts;
see \cite{KieSpohnCMP,SaffTotikBOOK}. 

\medskip\noindent
{\bf ACKNOWLEDGEMENT} 
This work was supported in parts through NSF Grant \# DMS-9623220.  



\end{document}